\documentstyle[11pt,aaspp4,epsf,rotate]{article}

\def\ugr{\, \lower4pt \hbox{$\buildrel > \over \sim$} \, }
\def\ukl{\, \lower4pt \hbox{$\buildrel < \over \sim$} \, }

\begin{document}

\title{A New Approach to Statistics of Cosmological Gamma-Ray Bursts}
\author{M. B\"ottcher\altaffilmark{1,2} \& C. D. Dermer\altaffilmark{2}}
\altaffiltext{1}{Rice University, Space Physics and Astronomy Department, 
MS 108\\
6100 S. Main Street, Houston, TX 77005 -- 1892, USA}
\altaffiltext{2}{E. O. Hulburt Center for Space Research, Code 7653, \\
Naval Research Laboratory, Washington, D. C. 20375-5352}

\bigskip
\centerline{\it Submitted to the Astrophysical Journal}
\bigskip

\begin{abstract}
We use a new method of analysis to determine parameters
of cosmological gamma-ray bursts (GRBs), assuming that their
distribution follows the star-formation history of the
universe. Spectral evolution is calculated from an
external shock model for fireball/blast wave evolution, 
and used to evaluate the measured
peak flux, duration, and $\nu F_\nu$ peak photon energy for a GRB
source occuring at a given redshift and with given values of
total energy, baryon-loading and environmental parameters. We
then fit model distributions of GRB sources to the observed
peak flux, duration and $\nu F_\nu$ peak photon energy 
distributions. We find that the observed width of the $E_p$
and duration distributions can not be explained by cosmological
redshift and time dilation effects. Rather, broad distributions
of total blastwave energies and bulk Lorentz factors are necessary
to explain the observed distributions simultaneously within the
framework of our unifying GRB model. We discuss implications
for source parameter distributions and determine a range
of burst parameters consistent with the data.

\end{abstract}


\section{Introduction}

The extragalactic origin of GRBs has been confirmed with the discovery 
of X-ray, optical and radio  afterglows of GRBs as a result of the
Italian-Dutch Beppo-SAX mission (e.g., Costa et al.\ \markcite{cea97}1997; 
van Paradijs et al.\ \markcite{vea97}1997; Frail \markcite{Frail98}1998).
At present, there are 8 GRBs with redshifts obtained from emission line 
measurements of host galaxy counterparts. These are GRB~980425 at redshift 
$z = 0.0084$ (\cite{kul98a}), GRB~970228 at $z = 0.695$ (\cite{djo99a}), 
GRB~970508 at $z = 0.835$ (\cite{bea98}), GRB~970828 at $z = 0.958$ 
(\cite{djo99}), GRB~980703 at $z = 0.966$ (\cite{djo98}), GRB~980613 
at $z = 1.096$ (\cite{djo99b}), GRB~990123 at $z = 1.60$ 
(\cite{kel99}) and GRB~971214 at $z = 3.418$ (\cite{kul98b}). 
The redshift inferred for GRB 980425 depends on the validity 
of the GRB~980425/SN1998bw association (\cite{gea98}). The  
host galaxy redshift of GRB~970508 supports the original 
redshift report obtained through absorption line measurements 
in its fading optical counterpart (\cite{met97}), and is 
strengthened by the recent Beppo-SAX announcement (\cite{piro99}) of a 
redshifted iron fluorescence feature found in the spectrum of its fading 
X-ray counterpart between $\approx 2.5\times 10^4$ and $6\times 10^4$ s 
following GRB 970508. 

It is not possible to construct a reliable GRB redshift distribution
from the statistics of eight GRBs.  Except for GRB 980425, however, 
whose nature remains controversial, the obtained redshifts place 
most GRBs at the cosmological epoch of active star formation. 
Several models for the origin of GRBs, such as the collapsar/hypernova 
scenario (Woosley \markcite{woo93}1993; Paczy\'nski 
\markcite{pac98}1998) or the supranova scenario of  Vietri \&
Stella (\markcite{vs98}1998), suggest that GRBs are physically related to
recent star formation. Their cosmological distribution should therefore
trace the star formation history of the universe (\cite{totani97a}). An
origin of GRBs involving stellar collapse events is also in accord 
with the small measured offsets of fading transient GRB counterparts 
with respect to the disks of the candidate host galaxies (\cite{kul98},
\cite{bloom99}).  By contrast, a larger offset is expected in 
the compact object coalescence scenario (e.g., Eichler et al.\ 
\markcite{eea89}1989; Narayan, Paczy\'nski, \& Piran \markcite{mpp92}1992) 
and, moreover, the redshift distribution of GRBs in this scenario 
is not required to follow the star formation history of the universe 
due to the large time delays between the  formation and merging of 
compact object binaries.

Several authors have calculated the GRB peak flux distribution,
testing the assumption of a GRB rate proportional to the 
observed star formation rate as given, e. g., by Lilly et 
al. (\markcite{lilly96}1996) and Madau et al. (\markcite{madau96}1996). 
Krumholz et al.\ (\markcite{kru98}1998) demonstrated that besides
a GRB spatial distribution following the star formation rate of
the universe, a variety of other luminosity and redshift 
distributions of cosmological GRBs are consistent with the 
observed peak flux distribution. Totani (\markcite{totani99}1999)
used an empirical correlation between the observed time-integrated
spectral shape of GRBs and their peak flux, and finds that 
the resulting peak flux distribution is only consistent with
the star formation history of  the universe if the star formation
rate between $z = 0$ and 1 is much flatter ($SFR [z = 1] \sim 4
\times SFR [z = 0]$) than deduced from UV observations ($SFR [z = 1]
\sim 15 \times SFR [z = 0]$), the former evolution being in agreement 
with independent estimates on the basis of galaxy evolution models 
(\cite{totani97}). He finds similar results for models invoking
binary neutron star or neutron star -- black-hole mergers. In 
contrast, Wijers et al.\ (\markcite{wij98}1998), assuming a 
standard-candle luminosity of GRBs, find reasonable agreement 
between their theoretical peak flux distribution and the one 
deduced from BATSE/PVO observations. 

This controversy indicates that the peak flux distribution 
data alone do not sufficiently constrain GRB parameters, 
in particular their luminosity and redshift distributions. 
Recently, Kommers et al.\ (\markcite{kom98}1998) have 
included very faint, ``non-triggered'' BATSE GRBs in a 
peak flux distribution study and demonstrated that even 
this improved data set cannot confidently distinguish
between a GRB rate tracing the star formation rate
and one tracing the redshift distribution of AGNs.

Previous theoretical GRB peak flux distribution studies have 
either used simple, non-evolving representations of the 
intrinsic burst spectra, such as a thermal bremsstrahlung
spectrum (\cite{fen93}) or a single power-law (\cite{kru98}),
or have summed over a sample of observed, time-integrated 
spectra (\cite{fb95}, \cite{wij98}, \cite{kom98}). Mallozzi 
et al.\ (\markcite{mal96}1996) demonstrated that the assumed 
intrinsic spectral shape of GRBs has a significant influence
on the results of theoretical GRB peak flux distribution studies.

The first attempt to combine peak flux distribution studies
with other statistical properties was done by Fenimore
\& Bloom (\markcite{fb98}1995), who deduced a typical
distance scale for GRBs from the observed effect of
time dilation on the burst duration by comparing the
brightest and dimmest BATSE bursts. They found that
in order to explain a time dilation by a factor of 2, the 
dimmest bursts had to be located at $z > 6$. However,
they pointed out that the implied isotropic burst 
luminosity was inconsistent with the observed peak flux
distribution, even in the case of a strong cosmological 
evolution of the comoving burst rate and of the luminosity
if both evolutions are parametrized as power-laws in 
$(1 + z)$. However, the significance of the observed 
peak flux -- duration correlation is highly controversial. 
Mitrofanov et al.\ (\markcite{mit93}1993) did not find any 
evidence for time dilation between dim and bright bursts, 
while Norris et al.\ (\markcite{nor94}1994) deduced a maximum 
redshift of $z \sim 2.25$ for the dimmest bursts from time 
dilation effects.  

Here we demonstrate the importance of considering
additional statistical properties to constrain the intrinsic
and environmental properties of GRB fireballs. These
include the duration distribution and the distribution of 
peak energies $E_p$ of the time-averaged $\nu F_\nu$ 
spectra of GRBs. The duration distribution of BATSE GRBs 
shows a pronounced bimodality (\cite{kou93}) between short
($t_{50} \lesssim 0.5$~s) and long ($t_{50} \gtrsim 3$~s)
bursts. This has been interpreted as evidence for two physically
distinct source populations by Katz \& Canel (\markcite{kc96}1996),
who found that the $\langle V / V_{max} \rangle$ distribution
of the populations of short and long bursts are significantly 
different from each other. Both populations show evidence for 
a cosmological origin ($\langle V / V_{max} \rangle < 0.5$), 
but the long bursts with $\langle V / V_{max} \rangle = 0.282 
\pm 0.014$ are located at larger distances and/or exhibit 
stronger cosmological evolution than the short bursts with 
$\langle V / V_{max} \rangle = 0.385 \pm 0.019$. This was
confirmed by Tavani (\markcite{tavani98}1998) who found
that the long / hard bursts show significantly stronger
deviations from a uniform distribution in Euclidean space
than other subclasses of GRBs. 

The $\nu F_\nu$ peak energies $E_p$ vary from burst to burst, 
although most are between $\approx 50$ keV and 1 MeV (Mallozzi 
et al.\ \markcite{mal97}1997; Strohmayer et al.\ 
\markcite{sea98}1998).  Assuming that the range of
$E_p$ intrinsic to the burst sources does not evolve with
redshift, Mallozzi et al.\ (\markcite{mal95}1995) found
that the $E_p$ vs.\ peak flux distribution of BATSE bursts
indicates a maximum redshift range of $(1 + z_1) / (1 + z_{100})
= 1.86^{+ 0.36}_{- 0.24}$ between bursts with peak fluxes of
$1$ and $100$~photons~cm$^{-2}$~s$^{-1}$, respectively. 

Previous work on modeling the peak flux distribution of
cosmological GRBs has mainly been carried out in a
model-independent way, using phenomenological
representations of the burst spectral and temporal
properties, without specifying physical parameters
of the burst sources. In particular, detailed predictions 
of theoretical blast-wave models for the peak power and 
temporal evolution of GRB spectra have never been used to
model the observed statistics of GRBs. In this paper, 
we use a parametric discription of the spectrum and 
spectral evolution predicted by the external shock model 
for cosmological GRBs (\cite{dcb99}) to calculate self-consistently 
the peak flux of a GRB at a given redshift, its duration $t_{50}$, 
and the observed peak energy $E_p$ of the $\nu F_{\nu}$ spectrum.  
Assuming that GRBs trace the star formation history of the
universe, this analysis helps to constrain not only the total 
energy of GRBs, but additional parameters such as the initial 
Lorentz factor (or baryon loading factor) $\Gamma_0$ of the 
GRB blast wave, the total energy $E_0$ deposited into the blast 
wave, the density of the circumburster material (CBM), and the 
equipartition factor $q$, which parametrizes the magnetic field 
strength and efficiency of electron acceleration in the blast 
wave.

In Section 2, we present the model equations for a GRB from a
fireball/blast wave which is energized, decelerates, and radiates 
by its interaction with a smooth CBM. We apply BATSE triggering 
criteria to this model in order to extract measured values of 
peak flux, duration and peak energy. A function describing the 
star formation history of the universe is given.  In Section 3, 
we calculate peak flux, $t_{50}$, and $E_p$ distributions 
for a GRB source population involving ranges of baryon loading 
factors and total fireball energies. Fits to the observed 
distributions are presented. The results are discussed in 
Section 4, and we summarize with implications for source 
models of GRBs.

\section{Peak Fluxes, Peak Energies, and Burst Durations}

Following the treatment given by Dermer et al.\ (\markcite{dcb99}1999a),
the photon number spectrum from a GRB located at redshift $z$ and its
temporal evolution is parametrized by the expression
\begin{equation}
\Phi (\epsilon, t; z) = {1 \over 4\pi d_L^2 \, \epsilon^2 \, m_e c^2}
\; { (1 + {\upsilon \over \delta}) \, P_p (t) \over [\epsilon /
\epsilon_p (t)]^{-\upsilon} + (\upsilon / \delta) \, [\epsilon
/ \epsilon_p (t)]^{\delta}},
\label{flux}
\end{equation}
where $\epsilon = E / (m_e c^2)$ is the dimensionless photon energy,
$d_L$ is the luminosity distance to the burst, $\upsilon$ and $\delta$
are the asymptotic low-energy and high-energy $\nu F_{\nu}$ slopes of 
the GRB spectrum, and $t$ is the time in the observer's frame. 
Throughout this study we use $\upsilon = 4/3$, while
a distribution of high-energy slopes $N(\delta) \propto \delta^{-0.5}$ 
for $\delta \ge 0.05$ is assumed (obviously, the normalization of the
spectral representation [\ref{flux}] diverges for $\delta \le 0$). This
distribution is in reasonable agreement with the distribution of 
GRB spectral indices with photon number indices $> 2$ determined in
the statistical analysis of high-energy  
spectra of BATSE GRBS by Preece et al. (\markcite{preece98a}1998a). The
measured distribution is only an approximation to the 
asymptotic high-energy 
distribution of slopes because the BATSE sensitivity range does not 
always sample photons with energies 
$\epsilon \gg \epsilon_p$.

The $\nu F_{\nu}$ 
peak photon energy is given by
\begin{equation}
\epsilon_p (t) = 3.0 \cdot 10^{-8} \, {q n_0^{1/2} \Gamma_0^4 
\over 1 + z}
\, \cases{ x^{-\eta/2} & for $0 \le x < 1$ \cr
           x^{-4g - \eta/2} & for $1 \le x \le \Gamma_0^{1/g}$. \cr}
\label{Ep}
\end{equation}
Here, $q = \sqrt{\xi_H \, (r/4)} \, \xi_e^2$ is the combined 
equipartition factor, containing the magnetic field equipartition
factor $\xi_H$, the electron equipartition factor $\xi_e$, and
the shock compression ratio $r$, and $\Gamma_0$ is the initial bulk
Lorentz factor or baryon loading factor of the blast wave. The term 
$\eta$ is the power-law index of the surrounding CBM density structure,
given by the smoothly varying function
\begin{equation}
n (x) = n_0 \, x^{-\eta}.
\label{n0}
\end{equation}
The CBM is assumed to be isotropically distributed about the source 
of the GRB. For simplicity, we let $\eta = 0$ in the calculations
presented here. The term $x = R/R_d$ is the radial coordinate $R$ 
in units  of the deceleration radius $R_d$, and is related to the 
observing time $t$ by 
\begin{equation}
x = \cases{ {t \over t_d} & for $0 \le x \le 1$ \cr
\left[ (2 g + 1) \, {t \over t_d} - 2 g \right]^{1/(2 g + 1)} & for
$1 \le x \le \Gamma_0^{1/g}$. \cr}
\label{xd}
\end{equation}
Here $t_d$ is the deceleration time scale of the blast wave in
the observer's frame, given by
\begin{equation}
t_d = {1 + z \over c \, \Gamma_0^{8/3}} \> \left[ {(3 - \eta) \, E_0
\over 4 \pi \, n_0 \, m_p \, c^2} \right]^{1/3}, 
\label{td}
\end{equation}
and $g$ parametrizes the radiative regime.  For a non-radiative
(adiabatic) blast wave, $g = 3/2 - \eta$, while $g = 3-\eta$ 
describes a fully radiative blast wave.  Throughout this study, 
we assume that the blast wave is uncollimated.  Beaming can be 
approximately implemented in the prompt and early afterglow phases
of a GRB by replacing the total energy $E_0$ with the quantity
$4\pi (\partial E_0/\partial \Omega)$, where $\partial E_0/\partial 
\Omega$ is the energy radiated per steradian into the solid angle 
element whose normal is directed within an angle $\lesssim 1/\Gamma_0$ 
of the direction to the observer. The peak $\nu L_{\nu}$ luminosity
$P_p (t)$ depends on the parameters of the model according to the 
relation
\begin{equation}
P_p (t) = {c (2 g - 3 + \eta) (4\pi m_pc^2)^{1/3}\over 2g \, (\upsilon^{-1}
+
\delta^{-1})\, (1 + z)^2}
\, n_0^{1/3} 
\, E_0^{2/3} \, \Gamma_0^{8/3} \, \cases{ x^{2 - \eta} & for $0 \le 
x < 1$ \cr
x^{2 - \eta - 4 g} & for $1 \le x \le \Gamma_0^{1/g}$. \cr}
\label{Pp}
\end{equation}

Using Eq. (\ref{flux}), we determine the peak flux $\Phi_p (z)$ of a 
GRB, averaged over the BATSE trigger time scale $\Delta t$ in the 
50 - 300~keV energy range, by scanning through $t_1$ and finding 
the maximum of the expression
\begin{equation}
\Phi_p (z) = \lower6pt\hbox{${\buildrel\max \over {t_1}}$} \>
\left\lbrace {1 \over \Delta t} \int\limits_{t_1}^{t_1 + \Delta t} dt 
\int\limits_{\epsilon_1}^{\epsilon_2} d\epsilon \> \Phi 
(\epsilon, t; z) \right\rbrace,
\label{peakflux}
\end{equation}
where $\epsilon_1 = 50/511$ and $\epsilon_2 = 300/511$. A burst 
producing a peak flux $\Phi$ has a probability $P_{\rm tr} 
(\Phi)$ to trigger BATSE. We parametrize the trigger efficiencies 
given by Fishman et al. (\markcite{fea94}1994) by
\begin{equation}
P_{\rm tr} (\Phi) = \exp[-(\Phi_0 / \Phi)^{\alpha}],
\label{Ptrig}
\end{equation}
where $\Phi_0$ and $\alpha$ depend on the trigger time scale: 

$$
\Phi_{0, 1024} = 0.26 \; {\rm photons \; cm}^{-2} \> {\rm s}^{-1},
\;\; \alpha_{1024} = 5.3 
$$
\begin{equation}
\Phi_{0, 256} = 0.53 \; {\rm photons \; cm}^{-2} \> {\rm s}^{-1},
\;\; \alpha_{256} = 5.7
\label{thresholds}
\end{equation}
$$
\Phi_{0, 64} = 1.05 \; {\rm photons \; cm}^{-2} \, {\rm s}^{-1},
\;\; \alpha_{64} = 5.5
$$
The different effective trigger thresholds are mainly
due to effects of variable background. In addition, the trigger
criterion corresponding to $\Delta t = 1024$~ms involves a 
selection bias toward bursts with durations $\gtrsim 1$~s. Due 
to this duration bias, shorter bursts with the same peak flux 
are less likely to trigger a GRB telescope because the total 
number of detected photons during the trigger time scale is 
smaller. Thus the effective integration time is determined
in this case by the burst duration rather than the
trigger time scale.

\begin{figure}
\rotate[r]{
\epsfysize=12cm
\epsffile[150 0 550 500]{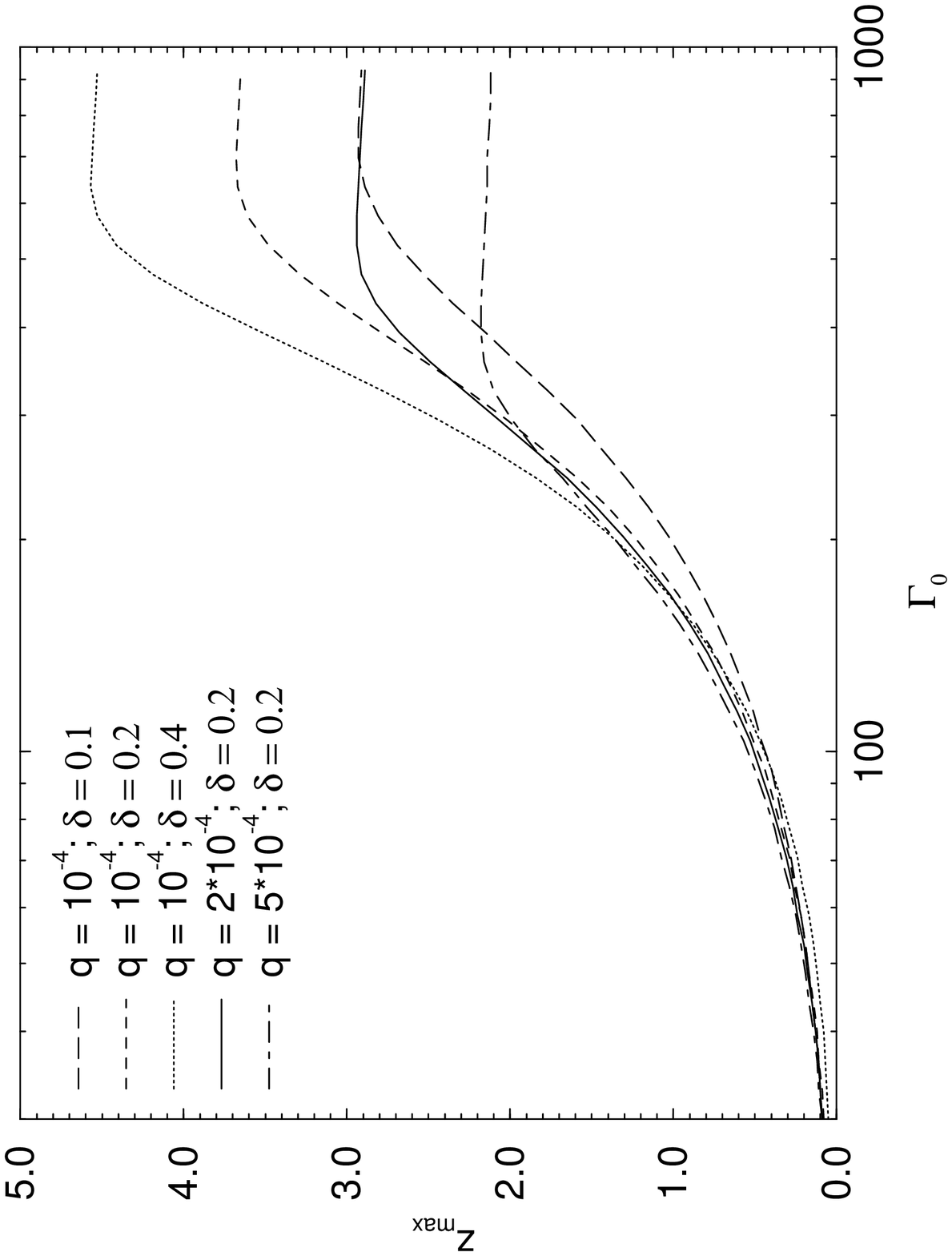}
}
\caption[]{The maximum redshift out to which a GRB can be detected 
above the 1024~ms trigger threshold $\Phi_{lim} \approx 
0.2$~photons~cm$^{-2}$~s$^{-1}$, as a function of baryon loading 
factor $\Gamma_0$, equipartition factor $q$, and high-energy
$\nu F_{\nu}$ spectral index $\delta$. Parameters: $E_0 = 
10^{53}$~erg, $n_0 = 100$~cm$^{-3}$, $g = 1.6$.}
\label{zmax}
\end{figure}

For a given peak flux value $\Phi_p$ and a given set of GRB parameters, 
an approximate limiting redshift $z_{\rm max} (\Phi_p)$ can be defined 
at which the GRB produces a peak flux $\Phi_p (z_{\rm max}) = \Phi_0$ 
corresponding to the respective trigger time scale. Fig.\ \ref{zmax} 
illustrates how $z_{\rm max}$ of a standard GRB varies with $\Gamma_0$, 
$q$, and $\delta$ for the 1024~ms trigger criterion. For fixed values 
of $q$, the peak flux increases with increasing $\Gamma_0$ until $E_p$ 
attains values greater than the photon energies of the detector triggering 
range.  At larger values of $\Gamma_0$, the emission recorded by 
a detector is dominated by the synchrotron emissivity spectrum 
$F_\nu \propto \nu ^{1/3}$ produced by a distribution of electrons 
with a low-energy cutoff. Because this portion of the spectrum is 
so hard, the peak flux, and therefore $z_{\rm max}$, declines at 
larger values of $\Gamma_0$, as shown in Fig.\ \ref{zmax}.

The Lorentz factor $\bar\Gamma_0$ of a fireball which produces the 
maximum peak flux in a detector sensitive to photons with energy 
$E_d$ is given by $\bar\Gamma_0 \approx 75 \, [(1+z)E_d/(m_e c^2 q 
n_0^{1/2})]^{1/4}$, using eq.\ (\ref{Ep}). The nominal BATSE triggering 
range, as noted above, is 50 keV $\leq E_d \leq$ 300 keV. For values 
of $\Gamma_0 \gtrsim \bar\Gamma_0$, $z_{max}$ is anticorrelated with 
$q$ because for larger $q$ the $\nu F_{\nu}$ peak is shifted towards
higher frequencies, causing the $(\Gamma_0, z_{max})$ curves to
level off at lower values of $\Gamma_0$. The maximum redshift is
also very sensitive to changes in $\delta$ for $\Gamma_0 \gtrsim
\bar\Gamma_0$. This is a consequence of the fact that a rapidly
increasing fraction of the radiated energy is emitted at $E > 300$~keV
as $\delta$ approaches 0. Obviously, $z_{max}$ is also strongly
dependent on the radiative regime $g$, determining the fraction
of the available energy which is converted into radiation. 

A differential peak flux distribution is calculated from the relation 
\begin{equation}
\Delta \dot N (\Phi_i < \Phi_p < \Phi_{i+1}) = {4 \pi \, c \over H_0} 
\int\limits_{z_{\rm max}(\Phi_{i+1})}^{z_{\rm max} (\Phi_i)} dz \> 
{\dot n_{GRB} (z) \, d_L^2 \, P_{\rm tr} (\Phi_p[z]) \over 
(1 + z)^3 \, \sqrt{(1 + \Omega_0 \, z) \, (1 + z)^2 - \Omega_{\Lambda}
\, (2 z + z^2)}}\; ,
\label{DN}
\end{equation}
(\cite{totani99}), where we choose $(\Omega_0, \Omega_{\Lambda})
= (0.3, 0.7)$ and $H_0 = 65$~km~s$^{-1}$~Mpc$^{-1}$. In this 
cosmology the luminosity distance $d_L$ is 
\begin{equation}
d_L = {c \over H_0} \, (1 + z) \int\limits_0^z {dz' \over 
\sqrt{(1 + \Omega_0 \, z') \, (1 + z')^2 - \Omega_{\Lambda} \, 
(2 z' + {z'}^2)}}.
\label{dL}
\end{equation}
Note that Eq. (\ref{DN}) refers to a mono-parametric distribution
of burst sources, for which a unique relation between peak flux and
redshift holds. In the lowest flux bin ($\Phi_0 < \Phi_p < \Phi_1$), 
the number of detected bursts will be determined by the trigger 
probability. Consequently, we use $z_{max} (\Phi_0) = \infty$.
$\dot n_{GRB} (z)$ is the comoving burst rate per comoving unit
volume. We assume that $\dot n_{GRB} (z)$ is proportional to the 
star formation rate (SFR), for which we use a simple representation 
of the function shown in Madau et al.\ (\markcite{mad98}1998):
\begin{equation}
SFR (z) [M_{\odot} \, {\rm yr}^{-1} \, {\rm Mpc}^{-3}] =
 \cases{10^{-3} \cdot 10^{z + 1} & for $z \le 1.1$ \cr
        0.13 & for $1.1 < z \le 2.8$ \cr
        4.2 \cdot 10^{-0.4 \, (z + 1)} & for $z > 2.8$. \cr}
\label{SFR}
\end{equation}
The $\nu F_{\nu}$ peak energy of the burst spectrum as a function
of redshift is evaluated at the time $t = t_d$  of peak power output 
of the GRB using eq.\ (\ref{Ep}). Thus $E_p/(m_ec^2) = 3.0\times 
10^{-8} qn_0^{1/2}\Gamma_0^4/ (1+z)$. 

\begin{figure}
\rotate[r]{
\epsfysize=12cm
\epsffile[150 0 550 500]{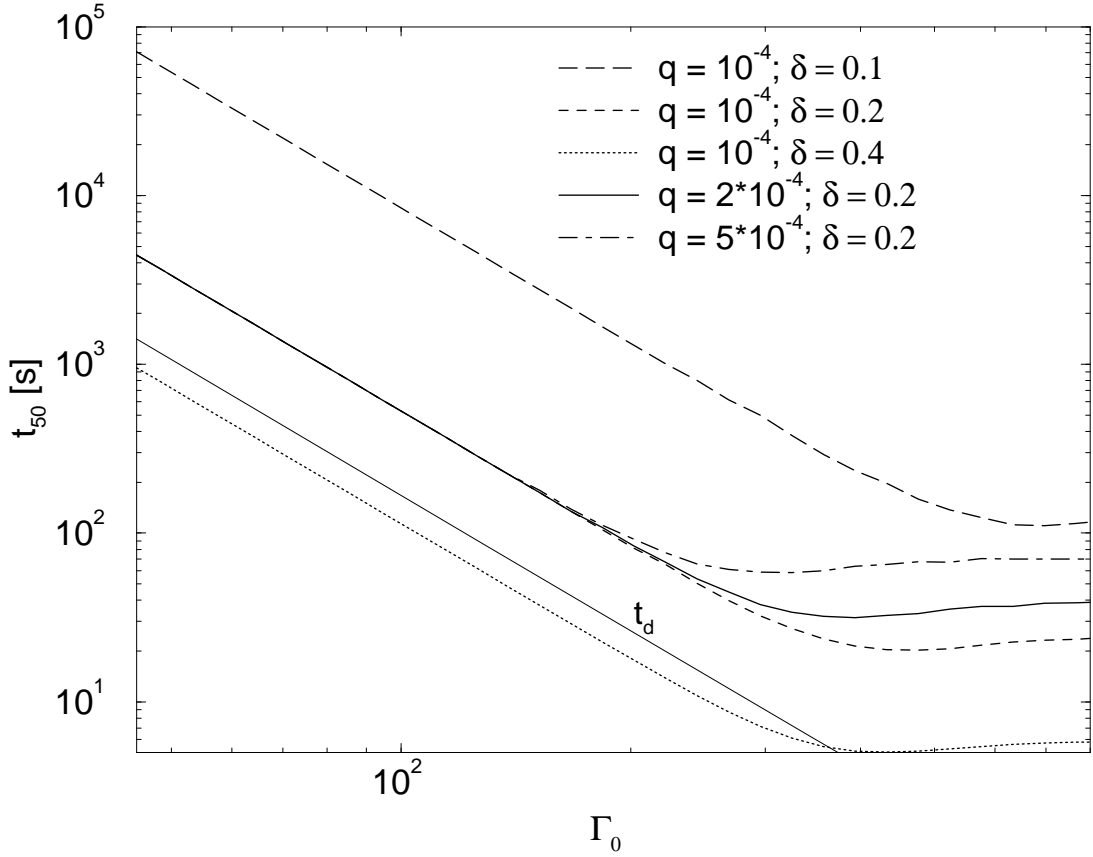}
}
\caption[]{Burst duration $t_{50}$ for a standard GRB located at $z = 1$,
as a function of $\Gamma_0$, $q$, and $\delta$. Other parameters: $E_0 = 
10^{53}$~erg, $n_0 = 100$~cm$^{-3}$, $g = 1.6$. The light solid curve 
shows the deceleration time $t_d$ for comparison.}
\label{t50}
\end{figure}

For the duration distribution, we calculate $t_{50}$ as the 
duration between the times when 25\% and  75\% of the total 
photon fluence in the 50 -- 300~keV band (from eq.\ [\ref{flux}]) 
has been received. In Fig. \ref{t50} we plot $t_{50}$ as a function 
of $\Gamma_0$, $q$, and $\delta$ for a standard GRB located at 
$z = 1$. For small values of $\Gamma_0$, for which $E_p$ is below 
or within the detector energy range, $t_{50}$ is proportional to 
the deceleration time $t_d \propto \Gamma_0^{-8/3}$. For larger 
$\Gamma_0$, $E_p$ is above the detector energy range, and the 
burst duration is determined by the time it takes for the 
$\nu F_{\nu}$ peak of the evolving burst spectrum to sweep 
through the detector energy range. For the parameters assumed 
here, $t_{50}$ is only weakly dependent on the initial bulk 
Lorentz factor in the high-$\Gamma_0$ limit. Using the asymptotic 
forms of eq.\ (\ref{flux}) and realizing that most photons are 
produced after $E_p (t)$ has swept through the detector energy 
range, one can analytically show that $t_{50} \propto t_d \; 
\Gamma_0^{(2g+1)/(g+\eta /8)} \propto \Gamma_0^{(1/g)-(2/3)}$ 
if $\Gamma_0 \gg \bar\Gamma_0$, where the last expression holds 
when $\eta = 0$. 

The burst duration $t_{50}$ is independent of the equipartition 
parameter $q$ for small values of $\Gamma_0$ and weakly positively 
correlated with $q$ for high $\Gamma_0$. There is a very strong
correlation between the burst duration and the high-energy spectral
index. As $\delta$ approaches 0, $t_{50}$ increases drastically
since the very hard tail of the spectrum produces a ``$\gamma$-ray 
afterglow'' which decays with time $\propto t^{-\chi}$, where
$\chi = [4 g (1 + \delta) - 2]/[2 g + 1]$ for $\eta = 0$. This
approaches a $t^{-1}$ decay if $\delta \to 0$ and $g \approx 1.5$.
The duration is also correlated with the radiative regime $g$
with $t_{50}$ decreasing by an approximately constant factor 
over the entire $\Gamma_0$ range considered here as $g$ increases.
For typical parameters, this downward shift can reach a factor of
$\sim 10$ as $g$ spans the range from 1.5 to 3.

The $E_p$ and $t_{50}$ distributions are calculated simultaneously
while scanning through redshift space according to Eq. (\ref{DN}). 
We caution that our analysis does not take into account any effects
due to the instrumental noise or the diffuse radiation backgrounds 
recorded by the BATSE detectors. The actual $t_{50}$ and $t_{90}$ 
durations of a GRB are expected to be slightly longer than measured 
because the additional background noise will dominate the emissions 
from a GRB at early and late times, particularly for weak GRBs. The 
actual $E_p$ distribution of those GRBs which trigger BATSE is, however,
not expected to be much different from the measured $E_p$ distribution,
because the background spectrum is subtracted in the spectral analyses 
which yield $E_p$ (see, e.g., Mallozzi et al.\ \markcite{mal95}1995). 

\section{Comparison with Observed Distributions}

We first applied the formalism described in the previous section 
to calculate peak flux, $E_p$, and duration distributions from 
cosmological GRBs with a single set of values for the burst parameters. 
The theoretical distributions were compared with the peak flux and 
duration distributions from the Third BATSE GRB catalog as compiled 
by Meegan et al.\ (\markcite{mee96}1996), and to the $E_p$ distribution 
of Mallozzi et al.\ (\markcite{mal97}1997). The values of $E_p$ were 
evaluated by Mallozzi et al.\  (\markcite{mal97}1997) by fitting 
the GRB spectrum from the 4B catalog (\cite{mee97}) with the 
Band function (\cite{band93}). We use the trigger time scale 
$\Delta t = 1024$~ms, corresponding to a peak flux threshold 
of $\Phi_{trig} \approx 0.2$~photons~cm$^{-2}$~s$^{-1}$. This 
is the trigger criterion used to extract the peak flux distribution 
data shown in Meegan et al. (\markcite{mee96}1996) and in our 
Figs. \ref{g_comp1}, \ref{g_comp2}, and \ref{fits}. We note that 
our analysis is properly applied to data sets which are produced 
by uniform triggering criteria, but that the $E_p$ distribution 
might be biased with respect to the duration and peak flux 
distributions since the $E_p$ data can be obtained only for 
the brighter GRBs. Moreover, the $E_p$ analysis uses 16 energy 
channel data, whereas the peak fluxes are based on the four 
energy channel discriminator data (R. Mallozzi, private 
communication, 1998). A uniform selection criterion for 
the peak flux, duration and $E_p$ distributions should be 
considered in future studies.

Figs.\ \ref{g_comp1} and \ref{g_comp2} show the derived 
distributions of GRB sources for mono-parametric models 
of the blast wave and surrounding medium.  The spread of 
measured values is thus a consequence of distance and 
cosmological effects and the GRB rate history, assumed 
throughout this paper to be proportional to the star 
formation rate SFR (Eq. [\ref{SFR}]). The dependence of 
the observables on the radiative index $g$ which parametrizes 
the radiative regime and thus determines the luminosity of the 
GRB, are shown in the individual figures. While the redshift 
and thus the peak flux distributions are very sensitive to 
slight changes in $g$, the $E_p$ and $t_{50}$ distributions 
are rather insensitive to the radiative regime. A comparison
of Fig. \ref{g_comp1} and \ref{g_comp2} demonstrates the
sensitivity of the resulting distributions on the total
energy $E_0 = 10^{52} \, E_{52}$~erg. The figures also show 
that the spread in $E_p$ and $t_{50}$ due to cosmological 
redshift alone is far less than the width of the observed 
distributions, indicating that they are probably dominated 
by a spread in the intrinsic GRB properties.

We note that our modeling result is not unique. There is 
at least one ambiguity in the sense that virtually 
indistinguishable peak flux, $E_p$ and duration distributions 
are obtained when varying the CBM density $n_0$ and the 
initial bulk Lorentz factor $\Gamma_0$, keeping the 
product $n_0 \, \Gamma_0^8$ constant. The reason for this 
ambiguity lies in the fact that the peak flux, $E_p$, and 
$t_d$ all depend only on the product $n_0 \, \Gamma_0^8$ 
(see eqs. [\ref{peakflux}], [\ref{Ep}], and [\ref{td}]). 
An average density of $n_0 \sim 10^2 - 10^5$~cm$^{-3}$ 
seems to be appropriate if GRBs are correlated with 
star-forming regions. 

As can be seen, the peak flux, duration, and $E_p$ distributions 
measured with BATSE requires fireballs with energies $\sim 10^{52}$
-- $10^{53}$~ergs, baryon-loading factors corresponding to $\Gamma_0
\sim 200$ -- $300$, and an equipartition factor $q\sim 10^{-3}$. 
The mean redshifts of such GRB sources lie typically near $z\sim 1$. 
The ability of the external shock model to account for the 
characteristic duration of GRBs was first noted by Rees and
M\'esz\'aros \markcite{rm92}(1992).

Fig.\ \ref{fits} shows the results of evaluating a large set 
($N \sim 2000$) of mono-parametric burst distributions and 
subsequently adding these distributions according to a variety 
of intrinsic parameter distributions. Apart from the high-energy 
slopes $\delta$ (see previous section), the total fireball 
energy $E_{52}$ and the bulk Lorentz factor $\Gamma_0$, 
are assumed to be distributed according to truncated 
power-laws, $N(E_{52}) \propto E_{52}^{-e}$, $N(\Gamma_0) 
\propto \Gamma_0^{-\gamma}$. The power-law indices $e$ 
and $\gamma$, and the boundaries $\Gamma_{0, {\rm min}}$, 
$\Gamma_{0, {\rm max}}$ are free parameters. Comparing 
the lower limits on $E_{52}$ deduced from observations 
of different GRBs, such as GRB~970508 (e.g., Waxman 
\markcite{waxman97}1997), GRB~980703 (Bloom et al.\ 
\markcite{bloom98b}1998b), and GRB~971214 (Kulkarni et al.\ 
\markcite{kul98b}1998b), not to mention GRB~980425 (Kulkarni et al.\ 
\markcite{kul98a}1998a), we know that the total available energy
in the blast wave may well vary from source to source by several
orders of magnitude. We assume that the values of $E_{52}$ vary
in the range $E_{52, {\rm min}} = 10^{-4}$ and $E_{52, {\rm max}} 
= 100$. We find that the resulting peak flux, $E_p$, and $t_{50}$
distributions are only weakly dependent on the actual values of 
these boundaries. Bursts with $E_{52} \lesssim 10^{-3}$ are only 
detectable at very small redshifts where the number density of
burst sources is assumed to be small due to the strong evolution 
of the star-formation rate (see eq. [\ref{SFR}]). They thus 
contribute very little to the observed distributions. For 
power-law indices $e \gtrsim 1$, the number of bursts with 
$E_{52} \gtrsim 100$ is strongly reduced compared to the 
lower-energy bursts, so that the contribution of bursts with 
energies beyond this value is of minor importance as well.

Dermer et al.\ (\markcite{dcb99}1999a, \markcite{dbc99}1999b)
argue that there could be a wide distribution in the baryon 
loading factor of GRB fireballs, of which preferentially those 
sources with $\nu F_{\nu}$ peaks between $\sim 50$~keV -- 
several MeV, corresponding to a baryon loading factor of 
$\Gamma_0 \approx \overline\Gamma_0$, are detected due to the 
triggering criteria of currently operating GRB monitors 
and the limitations of telescopes at X-ray and $\gg$~MeV 
energies. Detectability of burst sources with values 
of $\Gamma_0 \ll \overline\Gamma_0$ is strongly reduced 
because most of the flux will be emitted at energies below the 
sensitivity threshold of BATSE. The same is true, though to a 
lesser extent, for fireballs which produce blast waves with 
$\Gamma_0 \gg \overline\Gamma_0$. Thus we expect that the peak 
flux distribution observed from a population of burst sources 
with a variety of $\Gamma_0$ values will not strongly depend
on the actual range of baryon loading factors. However, due 
to the strong dependence of the peak energy $E_p$ on $\Gamma_0$ 
through eq.\ (\ref{Ep}), we expect that the observed $E_p$ 
distribution may well be influenced by the shape of the
$\Gamma_0$ distribution. The dependence of the burst duration 
on $\Gamma_0$ is weaker (see Fig. \ref{t50}), but still 
considerable. In terms of the deceleration time $t_d$, we 
have $t_d \propto \Gamma_0^{-8/3}$, so that a range in 
$\Gamma_0$ will also lead to a broadening of the $t_{50}$ 
distribution for a fixed value of $n_0$.

A reasonable match of our theoretical distributions simultaneously
with the peak flux, the $E_p$ and the $t_{50}$ distributions was 
achieved for $\Gamma_{0, {\rm max}} = 260$, $n_0 = 100$~cm$^{-3}$, 
$g = 1.7$, and $q = 10^{-3}$, $e \sim 1.5$, and $\gamma \sim 0$. 
A low value of $q \lesssim 10^{-3}$ is in agreement with the 
results of Chiang \& Dermer (\markcite{cd99}1999), who argue 
that such a small equipartition parameter is required in order 
to prevent rapid synchrotron cooling. A much larger value of 
$q$ would result in a low-energy synchrotron spectrum  $\Phi 
(\epsilon) \propto \epsilon^{-3/2}$, inconsistent with the 
observed hard low-energy asymptotes of BATSE GRB spectra 
(\cite{band93}, \cite{crider97}, \cite{preece98b}). To constrain
the intrinsic distributions in $E_{52}$ and $\Gamma_0$, we
fix $n_0$, $g$, $q$, and $\Gamma_{0, {\rm max}}$ to the values 
quoted above, and let the power-law indices $e$ and $\gamma$ 
vary as free parameters. In our fitting procedure we neglect
$\chi^2$ contributions from durations $t_{50} < 0.5$~s and
from $E_p$ values of $< 100$~keV because, as mentioned earlier, 
bursts with $E_p < 100$~keV tend to be too dim to allow the 
determination of spectral parameters such as $E_p$. We obtain 
a best fit simultaneously to all three distributions for 
$e = 1.52$ and $\gamma = 0.25$.
Fig. \ref{fits} shows the peak flux, $E_p$, $t_{50}$, and 
calculated redshift distributions from this burst 
population. Our fits are rather insensitive to the 
actual distribution of $\Gamma_0$ values in terms of  
$\Gamma_{0, {\rm min}}$ and the index $\gamma$ of the 
power-law distribution in $\Gamma_0$. GRBs with 
$\Gamma_0 \approx 220$ are preferentially detected 
for the value of $n_0 = 100$~cm$^{-3}$ chosen here, 
and the contribution of dirtier fireballs to the peak 
flux, $E_p$, and $t_{50}$ distribution is almost negligible.

\begin{figure}
\rotate[r]{
\epsfysize=12cm
\epsffile[150 50 570 560]{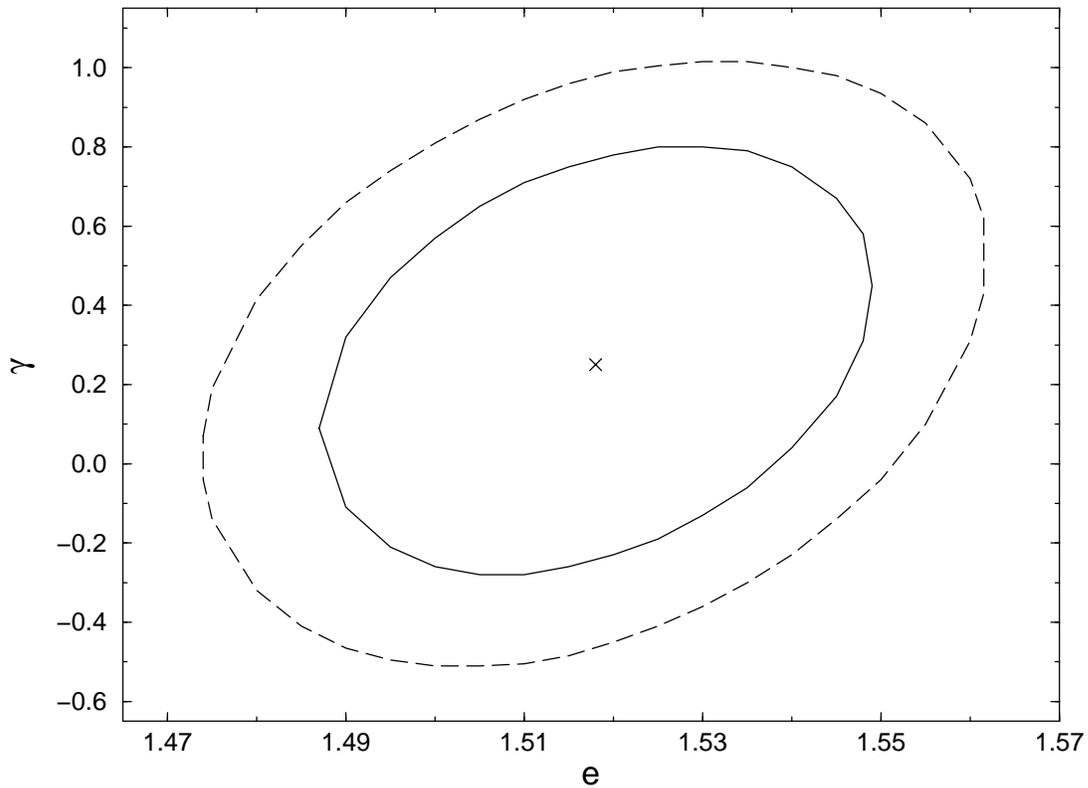}
}
\caption[]{$1 \sigma$ (solid line) and $2\sigma$ (dashed line) confidence
contours of the parameters $e$ and $\gamma$ parametrizing the distributions
of total blast wave energy and baryon loading factor, respectively
($N(E_{52}) \propto E_{52}^{-e}$, $N (\Gamma_0) \propto \Gamma_0^{-\gamma}$),
for a standard set of values: $g = 1.7$, $q = 10^{-3}$, $\Gamma_{0, {\rm max}}
= 260$, $n_0 = 100$~cm$^{-3}$.}
\label{contours}
\end{figure}

Fig. \ref{contours} shows the $1 \sigma$ and $2 \sigma$ 
confidence contours for the power-law indices $e$ and $\gamma$
of the $E_{52}$ and $\Gamma_0$ distributions, respectively, if the 
remaining relevant parameters are fixed to $g = 1.7$, $q = 10^{-3}$, 
$n_0 = 100$cm$^{-3}$, and $\Gamma_{0, {\rm max}} = 260$. The 
figure confirms that the actual shape of the distribution of 
baryon loading factors is only weakly constrained: A large range
of $\gamma$ values ($-0.2 \lesssim \gamma \lesssim 0.7$) yields 
acceptable fits to all three distributions. In contrast, the
power-law in $E_{52}$ is constrained to a narrow range of 
indices, $1.49 \lesssim e \lesssim 1.54$.

The normalization of our model peak flux distribution 
requires $\dot n_{\rm GRB} (z) = 4.43 \cdot 10^{-5} \, SFR  / 
M_{\odot}$, which yields a local GRB rate $\dot n_{\rm GRB} 
(z = 0) = 443$~GRBs~yr$^{-1}$~Gpc$^{-3}$. Assuming a local
galaxy number density of $4.8 \cdot 10^{-3}$~Mpc$^{-3}$
(\cite{wij98}), this is equivalent to a local GRB rate of 
92 Galactic events per Myr. This rate is a factor of 3700
higher than the result of Wijers et al.\ (\markcite{wij98}1998).
However, we note that the population assumed here contains many
undetected bursts for the existence of which, consequently, 
there is no evidence. 

The faintest detectable bursts in our model distributions
are located beyond a redshift of $z \gtrsim 4$. The maximum 
of the redshift distribution is located around $z \lesssim 1$, 
in good agreement with the redshifts from GRBs measured so 
far.  The small number of these GRBs does not yet allow an independent
statistical analysis. 

It is clear from Fig.\ \ref{fits} that the bimodality of the 
burst duration cannot be explained with the continuous burst 
source population assumed in our model calculations. The 
$t_{50}$ bimodality is not an instrumental effect,
but indicates that GRBs consist of at least 2 separate classes,
characterized by different physical parameters. The peak flux 
distribution of Meegan et al. (\markcite{meegan96}1996) was 
constructed on the basis of the 1024~ms trigger criterion, which 
misses most of the short bursts. Furthermore, all GRBs for which
redshifts could be measured as a consequence of their precise
localization by the BeppoSAX satellite, belong to the subclass
of long bursts. Thus, there is very limited information on the
short bursts so far. Therefore we do not attempt to model this
subclass of GRBs in this paper. With new data expected from 
the upcoming HETE~II mission, which might enable us to localize 
also several short bursts and determine their broadband spectral
characteristics and redshifts, a global statistical analysis 
similar to the one presented here might become possible.

The external shock model qualitatively explains the trend (Tavani
\markcite{tavani98}1998) that GRBs with harder spectra have, on 
average, smaller $\langle V/V_{\rm max} \rangle$ values.  Consider 
the simplification where fireballs are characterized only by 
$\Gamma_0$ and $E_0$. Blast waves with larger values of $\Gamma_0$ 
decelerate and radiate more rapidly and emit the bulk of their 
radiation at shorter wavelengths. They are therefore more luminous 
(see Fig.\ \ref{zmax}), shorter and harder. More energetic GRBs have
longer durations (eq. [\ref{td}]). Tavani's long/hard class of GRBs
therefore corresponds to the class of fireballs with the largest values of
$\Gamma_0$ and $E_0$, and these are the ones that can be detected from the
greatest distances and therefore exhibit the strongest non-Euclidean
effects.  By contrast, the long/soft class of GRBs represents less luminous
fireballs with smaller mean values of $\Gamma_0$ and $E_0$.  Only being
detectable from relatively nearby sites, these would display smaller
cosmological effects on the peak flux distribution.  Analogous reasoning
would also apply to the short class of GRBs, though the bimodality of this
population indicates that other parameters must also be taken into account.

We note that the external synchrotron shock model, as parametrized by eq.\ 
(\ref{flux}), specifically applies to the so-called ``Fast-Rise, 
Exponential Decay" FRED-like burst light curves (see Fishman 
\& Meegan \markcite{fm95}1995; \cite{dbc99}). While the overall 
spectral shape and luminosity (which determines the peak flux 
and $E_p$ distributions) may be similar for more complicated, 
spiky light curves --- which could arise from the interaction 
of the blast wave with an inhomogeneous medium (\cite{dm99}) 
--- the observed durations might be altered by this effect.
However, the bimodality of the duration distribution is not 
reasonably explained by a variety of light curves, and probably 
indicates a bimodality in parameter space (\cite{kc96}, 
\cite{tavani98}). 

\section{Summary and Conclusions}

We have used the cosmological blast wave model to fit 
simultaneously the peak flux, $E_p$, and $t_{50}$ distributions
as observed by BATSE, using an analytical representation of the
spectral evolution predicted by the external shock model. The GRB 
source distribution is assumed to trace the star formation history 
of the universe. For our standard cosmology ($H_0 = 65$~km~s$^{-
1}$~Mpc$^{-1}$, $\Omega_0 = 0.3$, $\Omega_{\Lambda} = 0.7$), the 
peak flux and $E_p$ distributions, and the $t_{50}$ distribution
of the population of GRBs with durations $t_{50} \gtrsim 0.5$~s 
can be modeled with a source population characterized by 
$N (\Gamma_0) \propto \Gamma_0^{-0.25}$, $\Gamma_0 \le 260$, 
$N(E_{52}) \propto E_{52}^{-1.52}$, $10^{-4} \le E_{52} 
\le 100$, $g = 1.7$, and $q = 10^{-3}$, if the density 
of the circumburster material is assumed as $n_0 = 
100$~cm$^{-3}$. Equally good fits can be found for 
different values of the CBM density, if the product 
$n_0 \, \Gamma_0^8$ is held constant. Our results
are fairly sensitive to the radiative regime of the 
blast wave and to the value of the equipartition 
parameter. 

The widths of the observed statistical distributions
can not be explained solely by cosmological redshift and
time dilation effects, but require a broad distribution of
intrinsic parameters, in particular of the total blastwave
energy $E_{52}$ and the initial bulk Lorentz factor $\Gamma_0$. 
Assuming that these parameters are distributed according to
single power-laws, the $E_{52}$ distribution is well constrained
for a fixed set of parameters which gives a good fit to the distributions, 
while the modeling results
are rather insensitive to the detailed shape of the $\Gamma_0$ 
distribution.

Our model calculation implies a local GRB rate of 
$443$~GRBs~yr$^{-1}$~Gpc$^{-3}$ or 92 Galactic events 
per Myr. This is a higher rate than obtained in some 
recent estimates (e.g., Wijers et al.\ 
\markcite{wij98}1998) and indicates a better
chance of detecting relatively young ($\lesssim 10^4$~yr)
GRB remnants in nearby galaxies. The photoionization signatures
of such young GRB remnants recently investigated in detail by 
Perna et al. (\markcite{perna99}1999) might then serve as a 
critical test for the assumed correlation between GRBs and 
star-forming regions.

The bimodality of the $t_{50}$ distribution can only be explained 
in the context of the present model if a bimodality in GRB source 
parameter space is assumed. 

\begin{figure}
\rotate[r]{
\epsfysize=12cm
\epsffile[150 50 570 560]{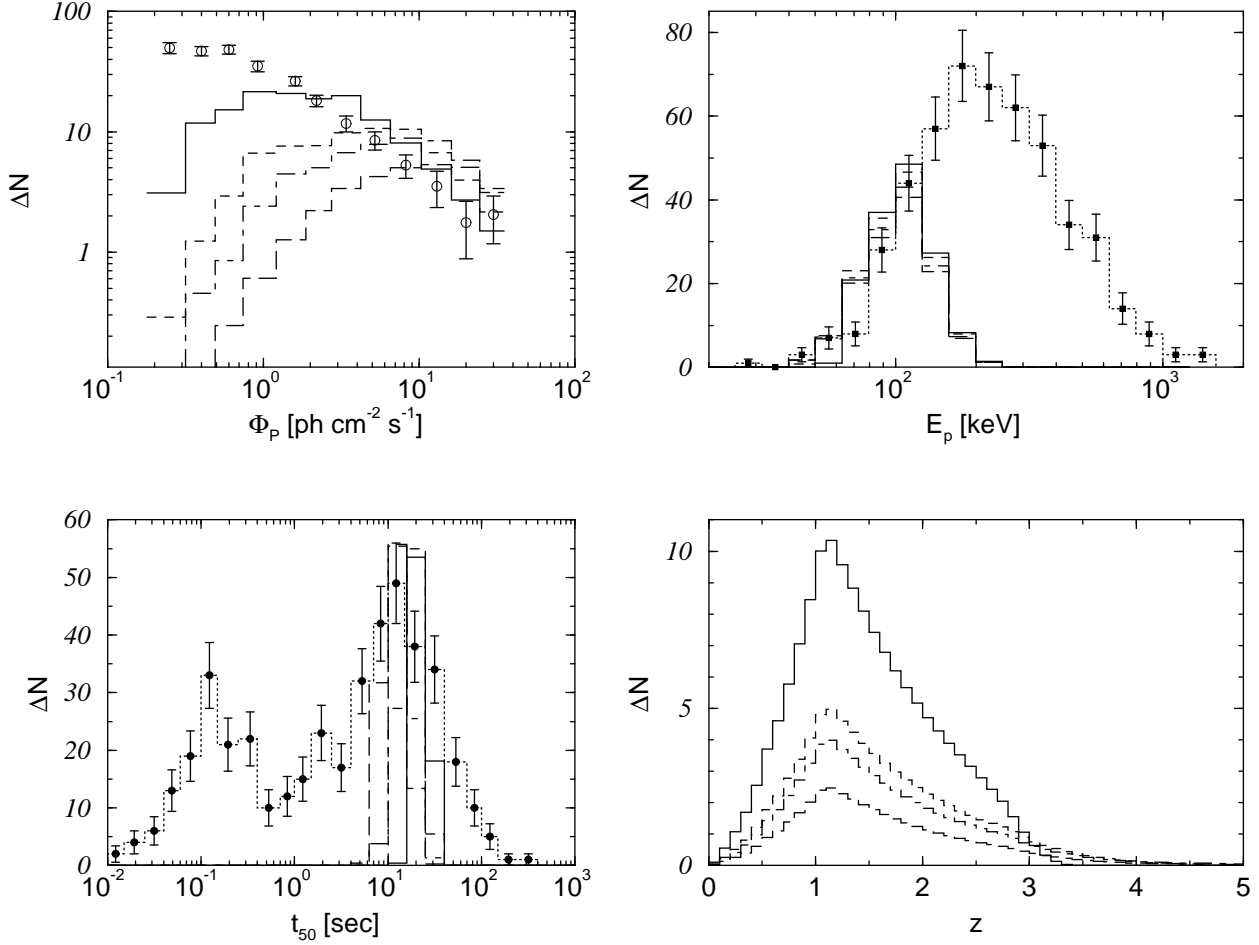}
}
\caption[]{Comparison of peak flux, $E_{pk}$, $t_{50}$ and redshift
distributions for a set of burst parameters $E_{52} = 50$, $\Gamma_0
= 240$, $n_0 = 100$~cm$^{-3}$, $q = 5 \cdot 10^{-4}$, for different 
radiative regimes of the blast wave, i. e. different values of $g$: 
$g = 1.6$ (solid histograms); $g = 1.8$ (dashed histograms); $g = 2.1$ 
(dot-dashed histograms); $g = 2.8$ (long-dashed histograms).}
\label{g_comp1}
\end{figure}

\begin{figure}
\rotate[r]{
\epsfysize=12cm
\epsffile[150 50 570 560]{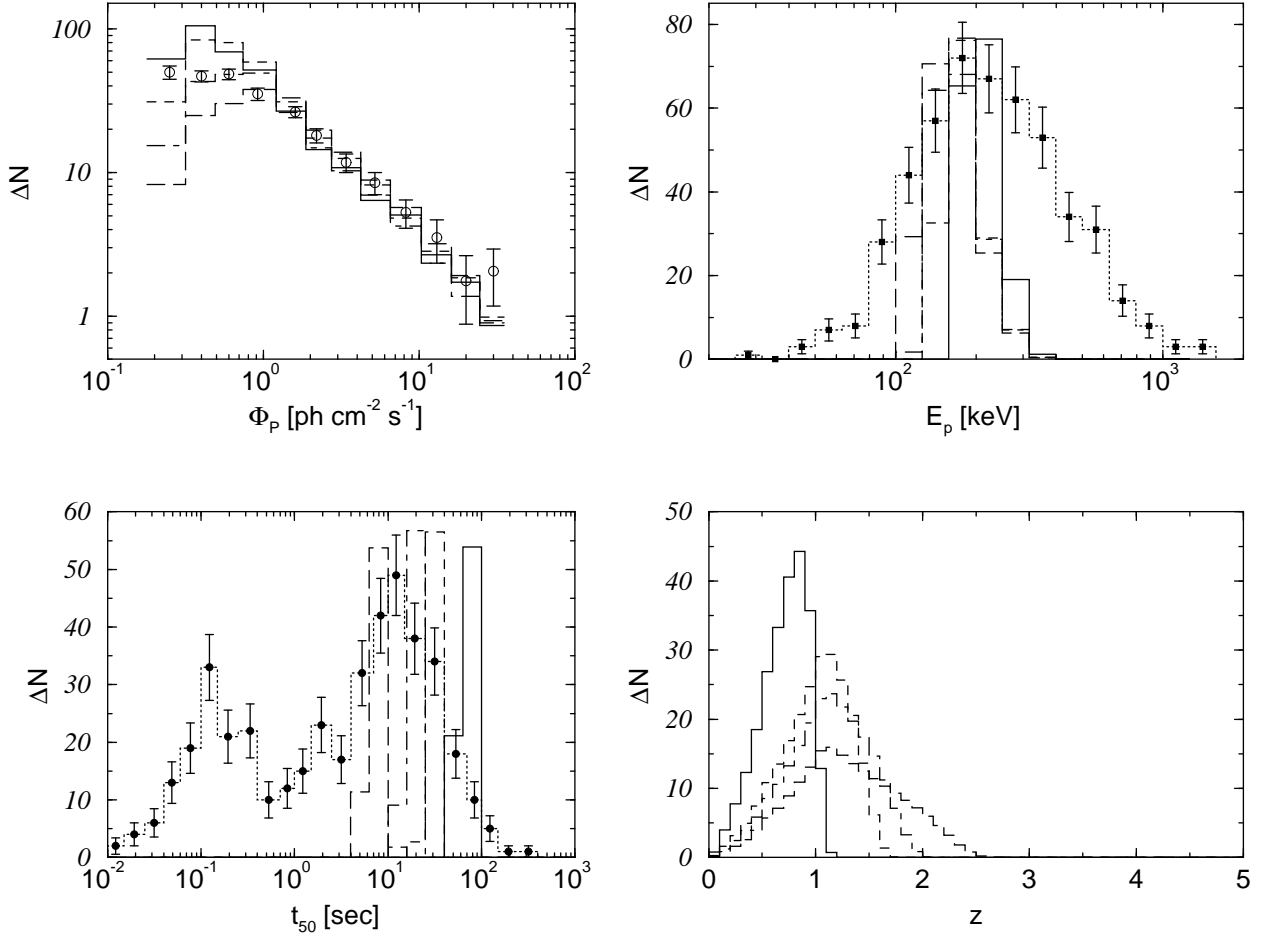}
}
\caption[]{Comparison of peak flux, $E_{pk}$, $t_{50}$ and redshift
distributions for a set of burst parameters $E_{52} = 3.8$, $\Gamma_0
= 220$, $n_0 = 100$~cm$^{-3}$, $q = 10^{-3}$, $\delta = 0.2$, for 
different values of $g$: 
$g = 1.6$ (solid histograms); $g = 1.8$ (dashed histograms); $g = 2.1$ 
(dot-dashed histograms); $g = 2.8$ (long-dashed histograms).}
\label{g_comp2}
\end{figure}

\begin{figure}
\rotate[r]{
\epsfysize=12cm
\epsffile[150 50 570 560]{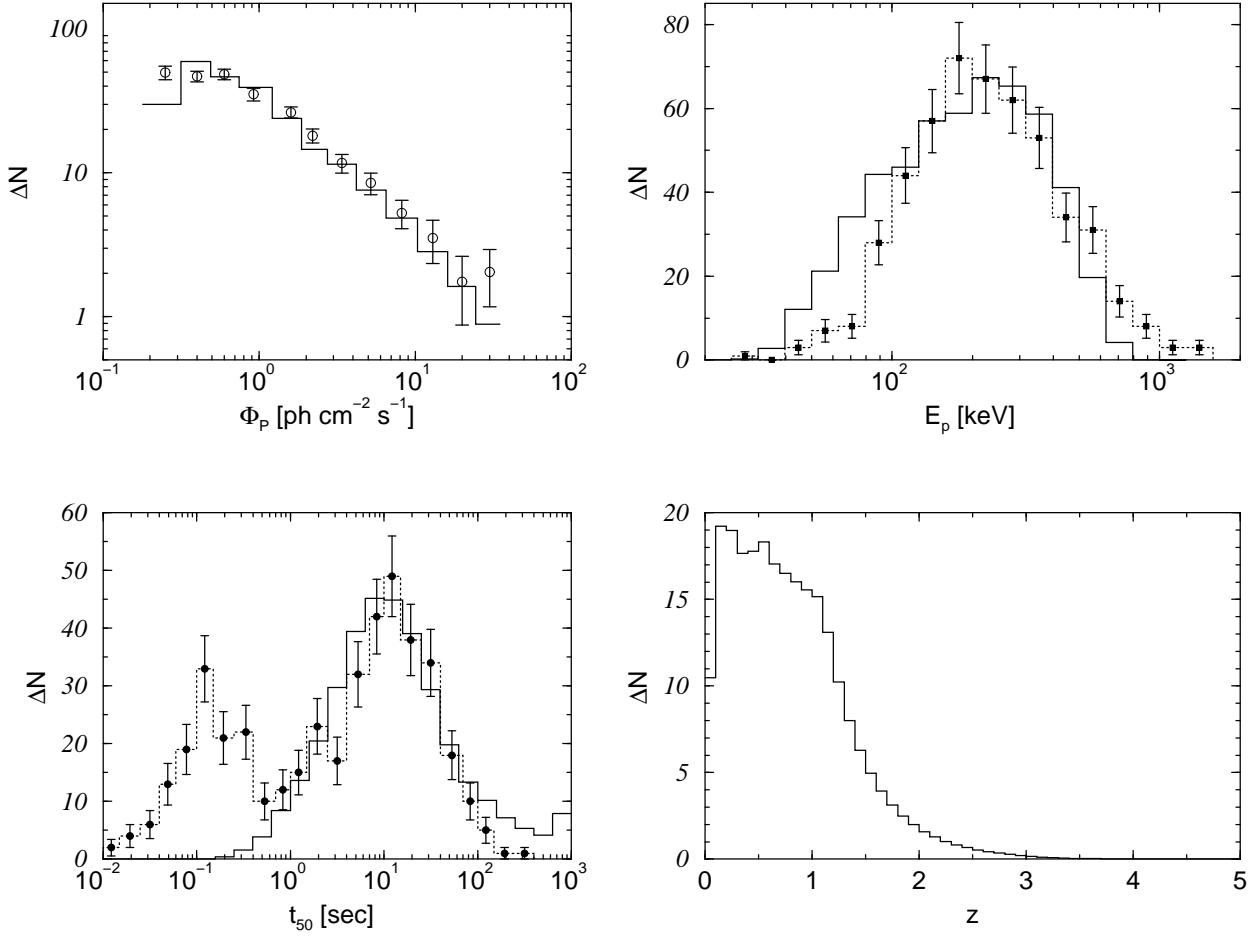}
}
\caption[]{Comparison of our model burst population to the 3B catalog 
peak flux, $E_{pk}$, and $t_{50}$ distributions, and model redshift
distribution. Parameters: $10^{-4} \le E_{52} \le 100$, $e = 1.52$, 
$\Gamma_0 \le 260$, $\gamma = 0.25$, $n_0 = 100$~cm$^{-3}$ $g = 1.7$, 
$q = 10^{-3}$.}
\label{fits}
\end{figure}

The simplified analytic form used to represent the observed
spectra from cosmological blast waves provides a reasonable 
representation of all three statistical distributions considered 
in this paper, thus lending strong support in favor of the external 
shock model of GRBs. The width of the observed $E_p$ distribution 
may be regarded as evidence for the existence of burst sources with 
a wide range of baryon loading factors $\Gamma_0$, as suggested by Dermer 
et al.\ (\markcite{dcb99}1999a, \markcite{dbc99}1999b). Our analysis 
only weakly constrains the existence of a  population of dirty fireballs 
with lower values of $\Gamma_0$, because these dirty fireballs are 
largely undetectable for BATSE  due to the rapid decline of the 
observed peak flux in the BATSE photon energy range with decreasing
$\Gamma_0$. BATSE is therefore insensitive to a dirty fireball 
population except for the few events that occur at redshifts $z \ll 1$. 
The small number of GRBs with $E_p \lesssim 100$~keV in the data
shown in Figs. \ref{g_comp1}, \ref{g_comp2}, and \ref{fits} might 
be, at least in part, an artifact of the selection bias mentioned 
at the beginning of the previous section, namely that $E_p$ can only 
be determined for bright bursts, while dirty fireballs, producing 
spectra with low $E_p$, are intrinsically dim.

Also the existence of a population of clean fireball burst 
sources cannot be constrained tightly on the basis of our 
results. However, this conclusion is strongly dependent 
on the radiative regime of the average GRB blast wave. 
For the value $g = 1.7$ found in our study to be an
appropriate choice to reproduce the peak flux, $E_p$ and
duration distributions, an additional population of clean
fireballs with $\Gamma_0$ much larger than the upper limit 
quoted above ($\Gamma_0^{max} n_0^{1/8}\approx 462$ cm$^{-3/8}$) 
can not be excluded, although it slightly worsens our 
modeling results. However, if a radiative regime $g \le 1.6$
is assumed, the decline of the peak flux with increasing
$\Gamma_0 > \overline\Gamma_0$ becomes much slower,
implying that clean fireballs would be detected as
high-$E_p$ bursts, inconsistent with the observed $E_p$
distribution. 

We therefore conclude that our analysis is in accord with
the existence of a class of dirty fireballs with $E_p \ll 
100$~keV which have generally not been detected with BATSE. 
Also the existence of a significantly larger fraction of cleaner 
fireballs than detected by BATSE cannot be excluded from
our analysis, although its consistency with observations
depends strongly on the best-fit radiative regime of the
average GRB blast wave.

Our results indicate that the most powerful GRBs can be detected
at redshifts $z \gtrsim 4$. The slope of $\sim 3/2$ in the peak
flux distribution seems to be a coincidence due to the particular
distribution of burst source energies, $E_{52}$, in combination
with the non-Euclidean space-time geometry. This is in accord
with the fact that the burst with the highest redshift measured
so far, GRB~971214, still has a rather high BATSE peak flux and
fluence, while the closest GRB for which a redshift estimate
exists so far, GRB~980425, was a rather dim burst. 

In summary, we use a parametrization of the blast wave model to 
calculate the peak flux, duration, and $\nu F_\nu$ peak energy 
of a GRB with a prescribed set of intrinsic and environmental 
parameters. We assume that the evolution of the GRB rate with 
redshift is proportional to the star-formation history of the
universe. By taking into account triggering properties of GRB 
detectors in our calculations, we make a detailed comparison of 
model distributions with BATSE results for the peak flux, $t_{50}$, 
and $E_p$ distributions. We have shown how these distributions 
can be self-consistently modeled and used to extract fireball 
model parameters. Our analysis shows that the observed
statistical distributions of observables of GRBs are in accord 
with the scenario that GRBs are associated with sites of active 
star formation, although alternative cosmological distributions
of GRB sources, which have not been considered in this paper,
cannot be excluded. Future work must consider evolutionary behaviors 
to test compact object coalescence scenarios.  We argue that the 
short, hard GRBs represent a separate population of burst sources.   

\acknowledgements{We thank the anonymous referee for very helpful
comments which led to considerable improvements of the manuscript.
We also thank J. Chiang for useful comments. This work is partially 
supported by NASA grant NAG~5-4055. CD acknowleges support from the 
Office of Naval Research.}

\newpage

\newpage

\newpage

\newpage

\newpage

\newpage

\begin{thebibliography}{}

\bibitem[Band et al.\ 1993]{band93}Band, D., Matteson, J., Ford, L., 
et al., 1993, ApJ, 413, 281

\bibitem[Bloom et al.\ 1998a]{bea98}Bloom, J. S., Djorgovski, S. G., 
Kulkarni, S. R., \& Frail, D. A. 1998a, ApJ, 507, L105

\bibitem[Bloom et al.\ 1998b]{bloom98b}Bloom, J. S., et al.\ 1998b, 
ApJ, 508, L21

\bibitem[Bloom et al. 1999]{bloom99}Bloom, J. S., et al., 1999, ApJL,
submitted

\bibitem[Chiang \& Dermer 1999]{cd99}Chiang, J., \& Dermer, C. D.,
1999, ApJ, 512, 699

\bibitem[Costa et al.\ 1997]{cea97}Costa, E., et al.\ 1997, Nature, 
387, 783

\bibitem[Crider et al.\ 1997]{crider97}Crider, A. W., Liang, E. P.,
Smith, I. A., et al., 1997, ApJ, 479, L39

\bibitem[Dermer et al.\ 1999a]{dcb99}Dermer, C. D., Chiang, J., \& 
B\"ottcher, M., 1999a, ApJ, 513, 656

\bibitem[Dermer et al.\ 1999b]{dbc99}Dermer, C. D., B\"ottcher, M., \&
Chiang, J., 1999b, ApJL, 515, L49

\bibitem[Dermer \& Mitman 1999]{dm99}Dermer, C. D., \& Mitman, K. E.,
1999, ApJ, 513, L5

\bibitem[Djorgovski et al.\ 1998]{djo98}Djorgovski, S. G., Kulkarni,
S. R., Bloom, J. S., Goodrich, R., Frail, D. A., Piro, L., \& Palazzi, E.,
1998, ApJ, 508, L17

\bibitem[Djorgovski et al.\ 1999a]{djo99a}Djorgovski, S. G., Kulkarni,
S. R., Bloom, \& Frail, D. A.
1999a, GCN 289

\bibitem[Djorgovski 1999]{djo99}Djorgovski, S. G., 1999, report at 
Santa Barbara Institute for Theoretical Physics Workshop on Gamma Ray Bursts

\bibitem[Djorgovski et al. 1999b]{djo99b}Djorgovski, S. G., et al., 1999b,
GCN Circ. 189

\bibitem[Eichler et al.\ 1989]{eea89}Eichler, D., Livio, M., Piran, T., 
\& Schramm, D. N. 1989, Nature, 340, 126

\bibitem[Fenimore et al.\ 1993]{fen93}Fenimore, E. E., Epstein, R. I.,
Ho, C., et al., 1993, Nature, 366, 40

\bibitem[Fenimore \& Bloom 1995]{fb95}Fenimore, E. E., \& Bloom, J. S.,
1995, ApJ, 453, 25

\bibitem[Fishman \& Meegan 1995]{fm95}Fishman, G. J., \& Meegan, C. A.,
1995, ARAA, 33, 415

\bibitem[Fishman et al.\ 1994]{fea94} Fishman, G. J., et al.\ 1994, 
ApJS, 92, 229

\bibitem[Frail 1998]{Frail98} Frail, D. 1998, in the Fourth Huntsville 
Gamma-Ray Burst Symposium, ed. C. A. Meegan, R. D. Preece, \& T. M. 
Koshut (AIP: New York), 563

\bibitem[Galama et al. 1998]{gea98}Galama, T. J., et al.\ 1998, 
Nature, 395, 670

\bibitem[Katz \& Canel 1996]{kc96}Katz, J. I., \& Canel, L. M., 1996,
ApJ, 471, 915

\bibitem[Kelson et al. 1999]{kel99}Kelson, D. D., Illingworth, G. I., 
Franx, M., Magee, D., \& van Dokkum, P. G., 1999, IAU Circ. 7096

\bibitem[Kommers et al.\ 1998]{kom98}Kommers, J. M., Lewin, W. H. G.,
Kouveliotou, C., et al., 1998, ApJ, submitted (astro-ph/9809300)

\bibitem[Kouveliotou et al.\ 1993]{kou93}Kouveliotou, C., et al., 1993,
ApJ, 413, L101

\bibitem[Krumholz et al.\ 1998]{kru98}Krumholz, M., Thorsett, S. E.,
\& Harrison, F. A., 1998, ApJ, 506, L81

\bibitem[Kulkarni 1998]{kul98}Kulkarni, S., presentation at ``Gamma-Ray
Bursts in the Afterglow Era," 3-6 November, 1998, Rome, Italy

\bibitem[Kulkarni et al.\ 1998a]{kul98a}Kulkarni, S., et al., 1998a,
Nature, 395, 663

\bibitem[Kulkarni et al.\ 1998b]{kul98b}Kulkarni, S., et al., 1998b,
Nature, 393, 35

\bibitem[Lilly et al.\ 1996]{lilly96}Lilly, S. J., LeF\`evre, O., 
Hammer, F., \& Crampton, D., 1996, ApJ, 460, L1

\bibitem[Madau et al.\ 1996]{madau96}Madau, P., Ferguson, H. C.,
Dickinson, M. E., et al., 1996, MNRAS, 283, 1388

\bibitem[Madau et al.\ 1998]{mad98}Madau, P., Pozzetti, L., \&
Dickinson, M., 1998, ApJ, 498, 106

\bibitem[Mallozzi et al.\ 1995]{mal95}Mallozzi, R. S., Paciesas, W. S.,
Pendleton, G. N., et al., 1995, ApJ, 454, 597

\bibitem[Mallozzi et al.\ 1996]{mal96}Mallozzi, R. S., Pendleton,
G. N., \& Paciesas, W. S., 1996, ApJ, 471, 636

\bibitem[Mallozzi et al.\ 1997]{mal97}Mallozzi, R. S., Pendleton, G. N.,
Paciesas, W. S., et al., 1997, in 4th Huntsville Symposium on Gamma-Ray
Bursts, eds. Meegan, C. A., Preece, R. D., \& Koshut, T. M., p. 273

\bibitem[Meegan et al.\ 1996]{mee96}Meegan, C. A., Pendleton, G. N.,
Briggs, M. S., et al., 1996, ApJS, 106, 65

\bibitem[Meegan et al.\ 1997]{mee97}Meegan, C. A., Paciesas, W. S., 
Pendleton, G. N., et al., 1997, in 4th Huntsville Symposium on Gamma-Ray
Bursts, eds. Meegan, C. A., Preece, R. D., \& Koshut, T. M., p. 3

\bibitem[Metzger et al.\ 1997]{met97}Metzger, M. R., Djorgovski, S. G.,
Kulkarni, S. R., et al., 1998, Nature, 387, 878

\bibitem[Mitrofanov et al.\ 1993]{mit93}Mitrofanov, I. G., et al., 1993, 
in 2nd Compton Symposium, ed. M. Friedlander, N. Gehrels, \& D. J. Macomb 
(New York: AIP Conf. Proc. 280), 761

\bibitem[Narayan et al.\ 1992]{npp92}Narayan, R., Paczy\'nski, B., \& 
Piran, T. 1992, ApJ, 395, L83 

\bibitem[Norris et al.\ 1994]{nor94}Norris, J. P., Nemiroff, R. J., 
Scargle, J. D., et al., 1994, ApJ, 424, 540

\bibitem[Paczy\'nski 1998]{pac98}Paczy\'nski, B., 1998, ApJ, 494, L45

\bibitem[Perna et al. 1999]{perna99}Perna, R., Raymond, J., \& Loeb, A.,
1999, ApJ, submitted (astro-ph/9904181)

\bibitem[Piro et al.\ 1999]{piro99}Piro, L., et al., 1999, ApJ, 514, L73

\bibitem[Preece et al.\ 1998a]{preece98a}Preece, R. D., Pendleton, G. N.,
Briggs, M. S., Mallozzi, R. S., Paciesas, W. S., Band, D. L., Matteson,
J. L., \& Meegan, c. A., 1998a, ApJ, 496, 849

\bibitem[Preece et al.\ 1998b]{preece98b}Preece, R. D., Briggs, M. S.,
Mallozzi, R. S., et al., 1998b, ApJ, 506, L23

\bibitem[Rees \& M\'esz\'aros 1992]{rm92}Rees, M. J., \& M\'esz\'aros, P.
1992, MNRAS, 258, 41P

\bibitem[Strohmayer et al.\ 1998]{sea98}Strohmayer, T. E., Fenimore, E. E., 
Murakami, T., \& Yoshida, A. 1998, ApJ, 500, 873

\bibitem[Tavani 1998]{tavani98}Tavani, M., 1998, ApJ, 497, L21

\bibitem[Totani 1997]{totani97a}Totani, T., 1997, ApJ, 486, L71

\bibitem[Totani et al. 1997]{totani97}Totani, T., Yoshii, Y., \& Sato,
K., 1997, ApJ, 483, L75

\bibitem[Totani 1999]{totani99}Totani, T., 1999, ApJ, 511, 41

\bibitem[van Paradijs et al.\ 1997]{vea97} van Paradijs, J., et al.\ 
1997, Nature, 386, 686

\bibitem[Vietri \& Stella 1998]{vs98}Vietri, M., \& Stella, L., 1998, ApJ,
507, L45

\bibitem[Waxman 1997]{waxman97}Waxman, E. 1997, ApJ, 489, L33

\bibitem[Wijers et al.\ 1998]{wij98}Wijers, R. A. M. J., et al.,
1998, MNRAS, 294, L13

\bibitem[Woosley 1993]{woo93}Woosley, S. E. 1993, ApJ, 405, 273

\end{thebibliography}
\end{document}